\documentclass[
prx,
 amsmath,amssymb,
 superscriptaddress,
 reprint,%
]{revtex4-2}

\usepackage{graphicx}% Include figure files
\usepackage{dcolumn}% Align table columns on decimal point
\usepackage{bm}% bold math
\usepackage{hyperref}% add hypertext capabilities
\usepackage[utf8]{inputenc}
\usepackage[T1]{fontenc}
\usepackage{etoolbox}
\usepackage{xcolor}

\usepackage{float}
\usepackage{amsfonts}
\usepackage{mathtools}
\usepackage[caption=false]{subfig}

\begin{document}
\begin{title}

%\preprint{AIP/123-QED}

\title{Josephson junction and inductor models in ADS}

\author{Ofer Naaman}

\date{\today}% It is always \today, today,

\begin{abstract}
This note is a follow up to Ref.~\cite{naaman2025modeling}, describing how to construct Josephson junction, inductor, and mutual inductance models using components that are available in the Keysight ADS core library. 
\end{abstract}

\maketitle
\end{title}

\section{Introduction}\label{sec:intro}
In Ref.~\cite{naaman2025modeling}, models for Josephson junctions, inductors, and mutual inductors were introduced. These models expose flux ports on these components to facilitate frequency domain simulations that enforce flux quantization conditions on the solution. Compiled behavioral models for these components were shipped with Keysight ADS QuantumPro starting with version 2025-U1. 

The goal of this note is to describe how to build similar models from basic components that are available with the ADS core library. We also aim to provide sufficient description of the models to allow users to implement them in other commercial or open-source circuit simulators.

\section{Models}\label{sec:models}
All models will be represented as subcircuit schematics with numbered and named ports. The components have parameters defined via the \texttt{File}$\rightarrow$\texttt{Design Parameters...} menu in the schematic editor, and these parameters and their description and units will be listed below. We assume that a global variable \texttt{Phi0} has been defined, as it is the case in version 2025 of ADS, and represents the numerical value of flux quantum \texttt{Phi0}$=2.068\times 10^{-15}$ when expressed in Weber units. The component symbol will be shown for reference, however these symbols are user-defined, so the users can create whatever artwork works for them. As in \cite{naaman2025modeling}, we use a convention that current flowing into the positive terminal of the component (inductor's dot, or junction's $+$ terminal) produces a positive flux across the flux port of the component.

\subsection{Josephson junctions}\label{sec:junctions}
Keysight ADS includes two behavioral Josephson junction models with its core library, \texttt{JJ2} and \texttt{JJ3}. The latter exposes a ground-referenced phase terminal that expresses the junction phase as a voltage, such that 1~V is equivalent to a phase of $2\pi$. The hard-coded ground reference for the phase terminal in \texttt{JJ3} is limiting the usefulness of the device, especially in circuits that involve compound SQUIDs or superconducting loops enclosing junction arrays. 

\begin{figure}[tbh]
\includegraphics[width=2.8in]{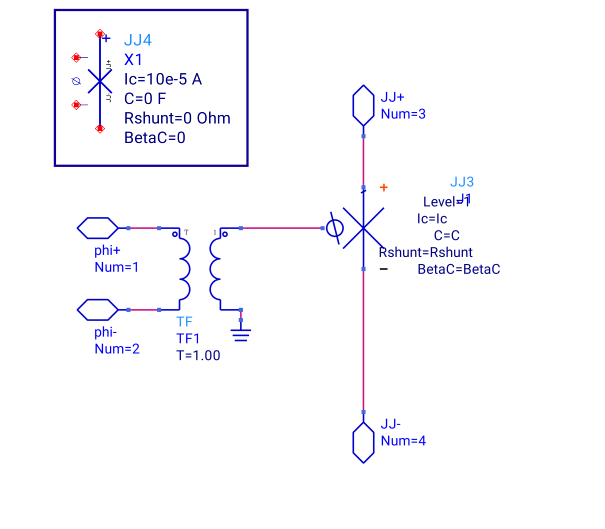}
\caption{\label{fig:JJ4} A junction model with balanced phase terminal, using an ideal transformer and a JJ3 component.}
\end{figure}

Starting with version 2025-U1, ADS QuantumPro bundle provides a behavioral component \texttt{JJ4}, which exposes the phase as a floating 2-terminal port. Fig.~\ref{fig:JJ4} shows how to construct a \texttt{JJ4}-equivalent device using only \texttt{JJ3} and an ideal 1:1 transformer that are available in core. The purpose of the transformer is to balance the phase terminal and provide a floating 2-terminal port. 

The parameters defined for the component mirror those that are defined for \texttt{JJ3} and are simply passed through to the \texttt{JJ3} instance in the schematic. These parameters are \texttt{Ic}, the junction critical current in units of Amperes; \texttt{C}, the junction capacitance in units of Farads; \texttt{Rshunt}, the junction shunt resistance in units of Ohms; and \texttt{BetaC}, the McCumber damping parameter (unitless).

\begin{figure}[tbh]
\includegraphics[width=\columnwidth]{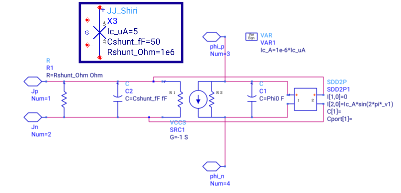}
\caption{\label{fig:JJ_shiri} A junction model following Ref.~\cite{shiri2023modeling}.}
\end{figure}

Users of other software packages could implement an equivalent model by following Ref.~\cite{shiri2023modeling}. Figure~\ref{fig:JJ_shiri} shows such a model, where compared to Ref.~\cite{shiri2023modeling}, we have included the junction shunt capacitance and resistance in the model, and avoided grounding the negative phase terminal. The model uses a Nonlinear Equation Based symbolically defined device (SDD), \texttt{SDD2P1} in the figure. The voltage at Port 1 (\texttt{\_v1}) of the SDD is the normalized phase across the junction $\delta/2\pi$, and the current at Port 2 implements the Josephson current equation $I=I_c\sin{\delta}$. The current in Port 1 is forced to zero (open circuit). The function of the voltage-controlled current source (\texttt{SRC1}) and capacitor \texttt{C1} are further illustrated in Sec.~\ref{sec:inductor}.

\subsection{Inductor}\label{sec:inductor}
\begin{figure}[tbh]
\includegraphics[width=\columnwidth]{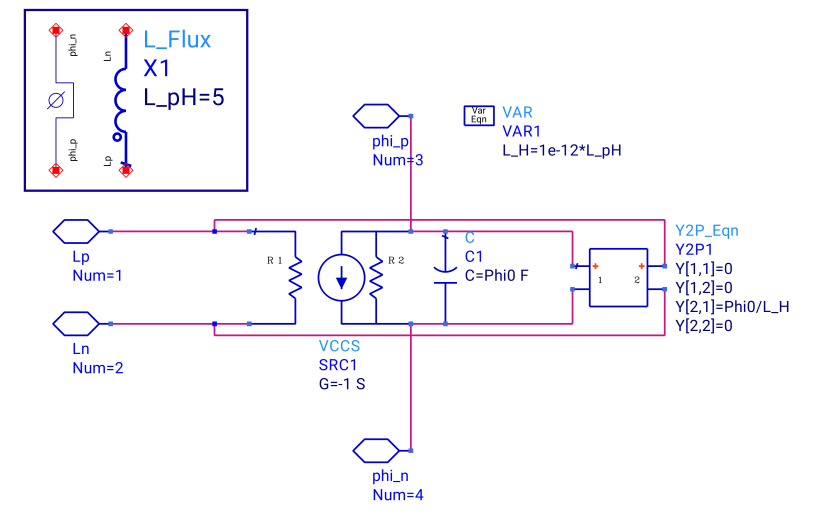}
\caption{\label{fig:L_Flux} A linear inductor with flux terminals, using a voltage-controlled current source, an integrating capacitor, and a 2-port Y-matrix Linear Equation-Based component.}
\end{figure}

Several inductor models were described in Ref.~\cite{naaman2025modeling}, including the \texttt{L\_Flux} model first shipped in ADS 2025. Fig.~\ref{fig:L_Flux} shows a schematic of an inductor with flux terminals, constructed with basic components available in ADS core. The only cell parameter is the unitless numerical value of the inductance, expressed in pH.

The voltage across the inductor terminals (Lp and Ln on the left), $V_L$, is converted to an internal current using a voltage controlled current source (VCCS, \texttt{SRC1} in the schematic) with transconductance of $G=-1$~S. This current is integrated using capacitor \texttt{C1} whose capacitance is set to \texttt{Phi0} Farads. That integral, appearing as the voltage on the capacitor is then representing the normalized flux across the inductor $V_\Phi=\frac{1}{\Phi_0}\int{V_Ldt}$, and is exposed on the phi\_p and phi\_n terminals. The voltage across the integrating capacitor (the `flux capacitor') is also fed to port 1 of a Linear Equation-Based block, \texttt{Y2P1} in the schematic, implementing a Y-matrix whose $Y_{21}$ element is $\Phi_0/L$ with all other elements set to zero (open circuit). Port 2 of the \texttt{Y2P1} block is connected back across the inductor terminal, driving the appropriate current through the inductor terminals, noting that the sense resistor $R1$ (and the shunt resistor $R2$) in \texttt{SRC1} is essentially an open circuit as set in the `properties' menu of the \texttt{SRC1} VCCS component.

\subsection{Mutual inductance}
A partial mutual inductance solution was presented in \cite{naaman2025modeling} and implemented in ADS 2025 QuantumPro package. That mutual had flux ports only on the secondary coil, and the details of the implementations were not given. 

\begin{figure*}[tbh]
\includegraphics[width=7in]{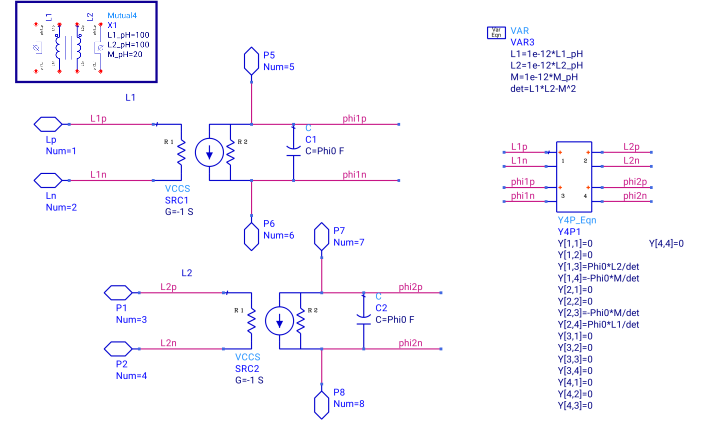}
\caption{\label{fig:Mutual4} A mutual inductance transformer component with 2 flux port. Nets in the figure are connected if they share the same name. Resistors $R_1$ and $R_2$ in both VCCS components are open circuit.}
\end{figure*}

Figure~\ref{fig:Mutual4} shows a model implementing two mutually coupled inductors with flux ports on both the primary and the secondary coil. In essence, this schematic expands on the inductor model of Fig.~\ref{fig:L_Flux}, and uses a $4\times 4$ Y-matrix instead of a $2\times 2$.

The parameters of the cell (all unitless numbers) are the numerical values of the primary inductor $L_1$, \texttt{L1\_pH}, the secondary inductor $L_2$, \texttt{L2\_pH}, and the mutual inductance $M$, \texttt{M\_pH}, all measured in pH. The variable equation block \texttt{VAR3} converts these parameters to Henry units, and calculates the determinant of the inductance matrix $\mathrm{det}=L_1L_2-M^2$. 

The voltages representing the flux on the primary inductor (nets phi1p and phi1n), and those on the secondary inductor (nets phi2p and phi2n) are connected to ports 3 and 4 of the linear equation based Y-matrix block \texttt{Y4P1}, respectively. The primary inductor terminals L1p and L1n, and those of the secondary inductor L2p and L2n, are connected to ports 1 and 2 of \texttt{Y4P1}, respectively. The Y-matrix block effectively calculates $\vec{I}=\mathbf{M}^{-1}\vec{\Phi}$, where $\vec{I}$ and $\vec{\Phi}$ are the vectors of currents and fluxes across the mutual transformer, and $\mathbf{M}$ is the inductance matrix:
\begin{align}
    I_1 &= \frac{1}{L_1L_2-M^2}\left(L_2\Phi_1-M\Phi_2\right)\\
    I_2 &= \frac{1}{L_1L_2-M^2}\left(L_1\Phi_2-M\Phi_1\right),
\end{align}
where $I_{1(2)}$ is the current in the primary $L_1$ (secondary $L_2$) inductor, and $\Phi_{1(2)}$ is the flux in the primary (secondary) inductor. Therefore, with the ports connected as described above, the Y-parameters in the schematic are set to
\begin{align}
    Y_{13}&=\Phi_0\frac{L_2}{L_1L_2-M^2}\\
    Y_{14}&=-\Phi_0\frac{M}{L_1L_2-M^2}\\
    Y_{23}&=-\Phi_0\frac{M}{L_1L_2-M^2}\\
    Y_{24}&=\Phi_0\frac{L_1}{L_1L_2-M^2}
\end{align}
and all other elements are set to zero (open circuit).

This cell can be pruned if only the secondary loop requires a flux port, for example, where the primary loop is intended to provide a flux bias from a non-superconducting circuit like a $50\,\Omega$ generator.

\section{Component test}
Components constructed following the schematics here, can be tested using some of the example circuits in Ref.~\cite{naaman2025modeling}. The mutual inductance component can be tested in a simple DC simulation to verify that the currents in either inductor produce the correct fluxes in both coils. Another useful test is an S-parameter simulation of a circuit containing two $LC$ resonators having the same frequency and coupled through the mutual inductance model\textemdash the simulated response should show the expected splitting in the resonator spectrum.

\section{Conclusion}
This note provides circuit schematics that should enable users of ADS, as well as other commercial or open-source EDA tools, to construct Josephson junction, inductance, and mutual inductance components with flux terminals. The components can facilitate frequency domain simulations of flux-quantizing superconducting circuits using the methods described in Ref.~\cite{naaman2025modeling}. 

%\clearpage

%\bibliography{references}

%\clearpage
\end{document}